\documentclass[aps,pra,epsf,twocolumn,amssymb,showpacs]{revtex4}
\usepackage{amsfonts}
\usepackage{amsmath}
\usepackage{graphicx}
\usepackage{eepic,epsfig,subfigure}
\usepackage{dcolumn}
\usepackage{bm,color,soul}
\usepackage[active]{srcltx}

\newcommand{\nn}{\nonumber}
\newcommand{\be}{\begin{equation}}
\newcommand{\ee}{\end{equation}}
\newcommand{\bea}{\begin{eqnarray}}
\newcommand{\ena}{\end{eqnarray}}
\begin{document}

\title{A proposed signature of Anderson localization and correlation-induced delocalization in an $N$-leg optical lattice}
\author{T.~A.~Sedrakyan$^{1,2}$, J.~P.~Kestner$^{1,2}$, and S.~Das Sarma$^{1,2}$}

\affiliation{Condensed Matter Theory Center$^1$ and Joint Quantum Institute$^2$,
Department of Physics, University of Maryland, College Park, MD 20742, USA}

\begin{abstract}
We propose a realization of the one-dimensional random dimer model and certain $N$-leg generalizations using cold atoms in an optical lattice.  We show that these models exhibit multiple delocalization energies that depend strongly on the symmetry properties of the corresponding Hamiltonian and we provide analytical and numerical results for the localization length as a function of energy.
We demonstrate that the $N$-leg systems possess similarities with their
1D ancestors but are demonstrably distinct.
The existence of critical delocalization energies leads to dips in
the momentum distribution which serve as a clear signal of the localization-delocalization transition.
These momentum distributions are different for models with different group symmetries and are identical for those with the same symmetry.
\end{abstract}

\maketitle
\section{Introduction}
Fifty years after Anderson's original prediction \cite{And58} of electron localization in weakly disordered solids, experiments demonstrated localization of
ultracold neutral atoms in a disordered optical potential \cite{Aspect08,Inguscio08}.  The great advantage of these atomic systems is the control of disorder
form and strength, interactions, dimensionality, etc., in principle allowing for unambiguous observation of the phenomenon under a variety of circumstances.
However, the localization observed so far is \emph{qualitatively} indistinguishable from classical localization of particles in a disordered trapping potential.  Identifying the observed localization as a quantum effect requires special care \cite{Aspect08}.
It is thus of utmost interest to use the flexibility of the atomic systems to engineer disorder such that a unique signature of Anderson localization emerges.
In this work we show both analytically and numerically that a general class of disorder models similar to the well-known random dimer model exhibits multiple
delocalization energies.  We propose an experimental realization of this class of models with ultracold atoms in optical lattices, using current state-of-the-art
techniques.  Such an experiment would feature an unusual and unmistakable experimental signature of quantum localization in the form of sharp dips in the measured
momentum distribution.  The possibility of the direct observation of the quantum localization-delocalization transition in suitably designed optical lattices, particularly in situations where there should be no such classical transitions, is the primary motivation of our theoretical work.

On another front, it is very interesting to investigate the effect of introducing correlations to the disorder potential in Anderson-type models in \emph{higher} dimensions.
It is well-known that Anderson localization in one dimension can be circumvented under certain conditions for disorder possessing long-range \cite{Moura98,Izrailev99,Garcia} or short-range \cite{DasSarma, SS3, DWP, 7SS, SS6, TS, TSO, Vignolo10} correlations or for pseudorandom models where the disorder arises from an additional potential incommensurate with the lattice potential \cite{R1,R2,R3,R4}.  One well-studied example of such one-dimensional delocalization is the one-dimensional (1D) random dimer model \cite{DWP}, first studied in the context of the existence of extended resonance states in incommensurate 1D lattices with complex unit cells \cite{DasSarma} and later extensively applied to the metal-insulator transition in conducting polymers \cite{7SS,SS6}.  In this 1D random dimer model, the localization length of the single-particle states diverges at a pair of critical energies.  Similar behavior has also been shown for a two-leg ladder version of the same model \cite{TSO}.  In this work, we consider several distinct models in an $N$-leg system.  Such systems fill an interesting niche between 1D and 2D delocalization behavior.  Observing correlation-induced transitions in this case using the clean and controllable optical lattice system is an intriguing prospect.  (The word "correlation" throughout this article and in the title of this paper refers exclusively to correlations in the disorder within the noninteracting quantum tight-binding model -- mutual interaction between the particles themselves, which is often also referred to as "correlations" in the literature, is beyond the scope of the current work.)

\begin{figure}[]
\subfigure[]{\includegraphics[width=.45\columnwidth]{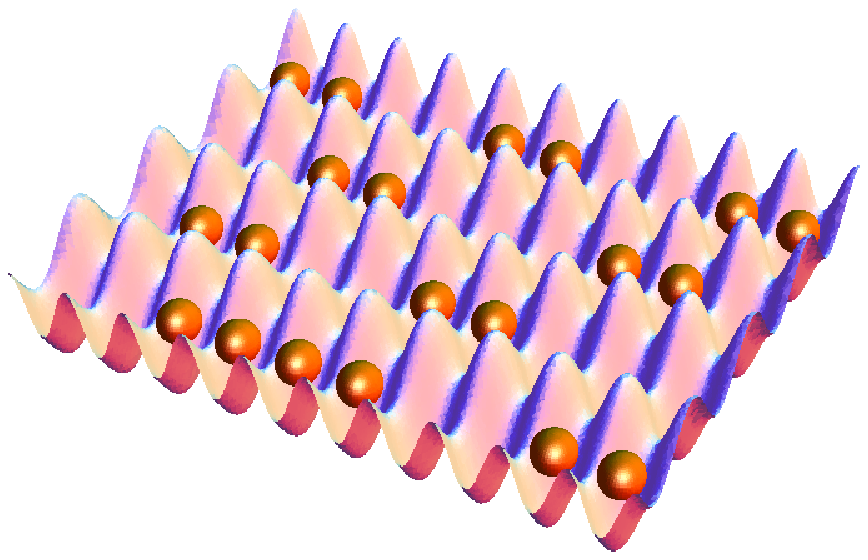}\label{fig:uncorr}}
\subfigure[]{\includegraphics[width=.45\columnwidth]{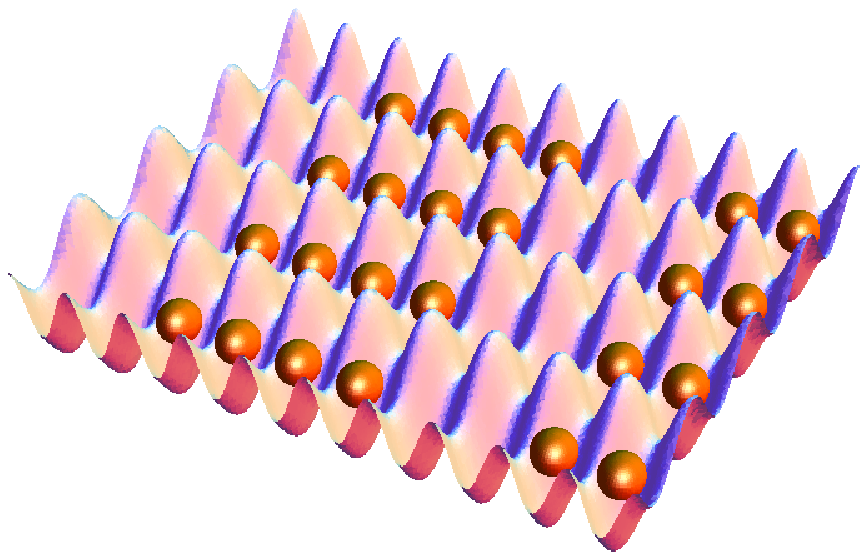}\label{fig:vertstripe}}
\subfigure[]{\includegraphics[width=.45\columnwidth]{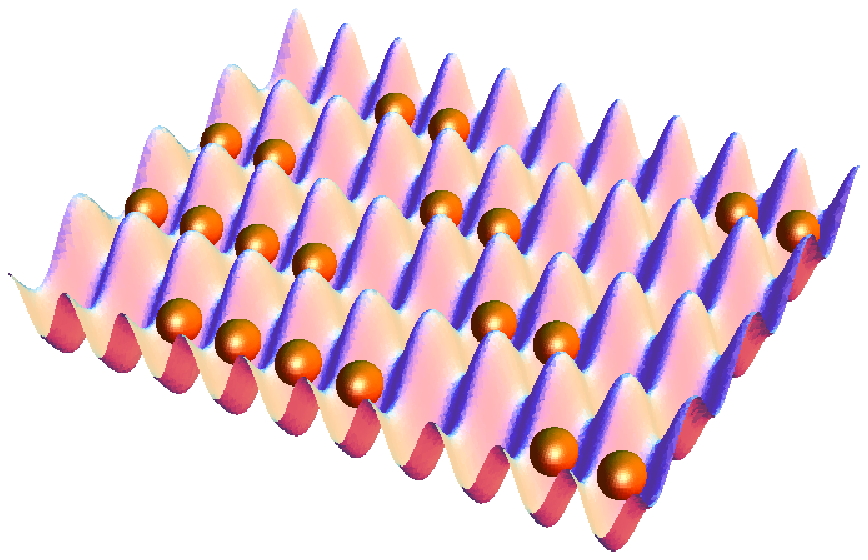}\label{fig:diagstripe}}
\subfigure[]{\includegraphics[width=.45\columnwidth]{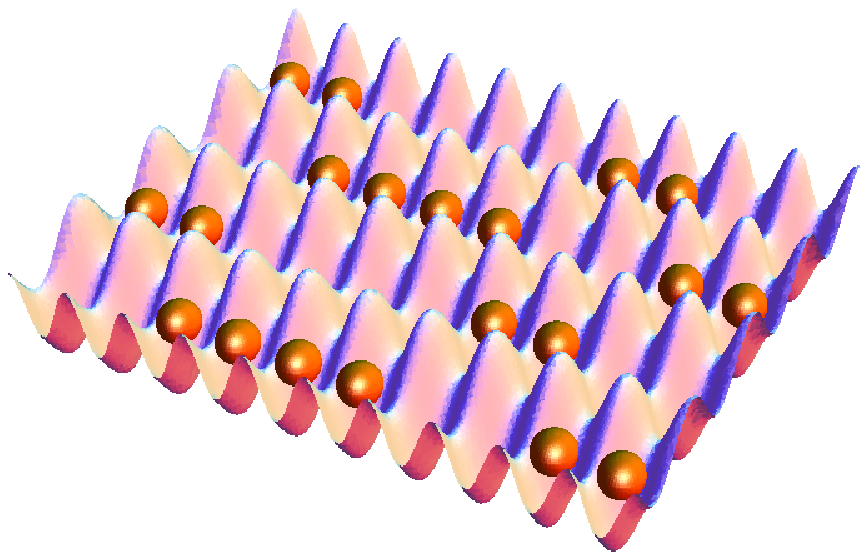}\label{fig:stagstripe}}
\caption{(Color online) Impurity pattern for a four-leg lattice with horizontal (i.e., intraleg) dimer correlations and the following vertical (i.e., interleg) correlations (a) uncorrelated model, (b) vertical stripe model, (c) diagonal stripe model, and (d) staggered stripe model.}\label{fig:stripes}
\end{figure}

When one couples a set of 1D chains, an additional degree of freedom and associated symmetries come into the picture. Depending on the particular experimental scheme, the disorder potential along the vertical (i.e., interleg) direction may respect different symmetries.  In the full 2D case, for instance, the potential may be invariant under the group of vertical lattice translations, $T_{vert}$.  One may then apply this potential on an $N$-leg lattice.  Below we call such a potential a vertical stripe model, see Fig.~\ref{fig:vertstripe}.
The vertical stripe scheme associated with the $N$-leg model with intraleg dimer correlations turns out to be trivial in the sense that it can be mapped onto $N$ separate dimer chains by a unitary transformation. This fact implies that the physical observables are
essentially the same as those for the 1D dimer model, which can also be inferred from the separability of the disorder potential.

It is clear that there is an infinite number of ways in which this separability can be broken.  To gain insight into which cases will preserve the existence of delocalization transitions and will take us out of the simple effectively 1D dimer model, one needs to consider the symmetries of the model in a general case.

In the present work we will show that breaking the translational invariance of the potential in the vertical direction and reducing the symmetry of the model down to $T_{vert}/{\cal G}$, where ${\cal G}$ is a discrete group, can lead to qualitatively new effects that can clearly distinguish the disordered $N$-leg model
from the 1D random dimer model.
We argue -- and show for particular examples -- that the physical phenomena in these schemes are determined by the properties of groups ${\cal G}$ for each given number of legs, $N$. In this way one effectively characterizes properties of correlated disordered models based on the factorization of the group of translations with respect to
discrete groups\cite{group}. This is one of the key results of our paper, and it is in some sense reminiscent of (and similar to) the classification of fully (uncorrelated) disordered systems \cite{Zirn,Cas} and topological phases \cite{Lud} based on the Cartan classification of symmetric spaces. This is important especially because it is known that the
type of correlations can have a strong effect on localization
properties of disordered systems. Therefore characterization of schemes according to their symmetry properties directly links these schemes with the expected universal behavior of observables that can be measured experimentally.  Our current work and the earlier works on the classification of uncorrelated disordered systems \cite{Zirn} and topological phases \cite{Lud} demonstrate that underlying group symmetries provide essentially a complete classification of the possible universality classes of noninteracting quantum models although the inclusion of interaction effects in these group classifications remain an important unsolved challenge.

In the models where the group ${\cal G}\equiv Z_2$ (we term this model a staggered vertical stripe model, see Fig.~\ref{fig:stagstripe}), the invariance under the group of translations $T_{vert}$ is broken along the vertical direction on $2$ lattice spacings, $\tau |n,i\rangle = |n, i+2\rangle$, where $\tau\in T_{vert}$.
This qualitatively changes the physical properties of the system compared to the vertical stripe model, where the translational invariance in the vertical direction holds
for all translations.
Breaking of the translational invariance  (i) reduces the symmetry of the model from $T_{vert}$ down to $T_{vert}/Z_2$; (ii) preserves the existence of truly
delocalized states in a model with $M$-mers,
as we will see from our numerical calculations of the localization length for the staggered vertical stripe model; and (iii) leads to qualitatively different forms of the momentum distributions for different number of legs, $N$, and correlations, $M$.

One could also generalize the staggered model and consider, for example, longer periodic correlations of the site energies, $\epsilon_{n,i}$, in the vertical
direction. For example one can violate the translational invariance in the vertical direction on three and more
($m$ in general) lattice spacings,
which will reduce the symmetry by the discrete group $Z_m$ down to the factor space $T_{vert}/Z_m $.
As in the vertical stripe model case, the presence of this symmetry is the origin of the appearance of delocalized states
in a corresponding $M$-mer ($M=2,3,..$) model.

  In this work we will discuss in detail and contrast two simple but spectacular cases out of the infinite number of ways in which the separability of the disorder potential can be broken. In the first example, which is termed the $N$-leg diagonal stripe model (see Fig.~\ref{fig:diagstripe}), the separability is broken but also the invariance under the group of translations $T_{vert}$ in the vertical direction is broken completely (as opposed to the case where it is broken down to $T_{vert}/Z_m $). We show that such a modification leads to the complete localization of states. The second example is the $N$-leg staggered vertical stripe model which clearly yields new experimentally observable effects which we discuss in detail.

Generally, violation of the translational invariance in the vertical direction by any discrete group, ${\cal G}$, will work in a similar way and one should expect the presence of delocalized states here as well. Amazingly, this property implies the classification of the correlated disordered systems characterized by the orientation-preserving transformations corresponding to ${\cal G}$.

The paper is organized as follows. First, we propose a procedure to experimentally realize the 1D random dimer model in an optical lattice and measure the delocalization energies in Sec.~\ref{sec:1D}.  This model offers
a unique opportunity to both explore correlation-induced delocalization and observe an unambiguous signal of Anderson localization.  Second, in Sec.~\ref{sec:Nleg} we generalize to
anisotropic short-range-correlated disorder in an $N$-leg ladder.  We analytically and numerically investigate localization properties and propose experimental
techniques to realize these models with ultracold atoms in an optical lattice. We conclude in Sec.~\ref{sec:conclusions} with a brief summary.  Exact solution of the vertical stripe model and some technical details are given in the Appendices.

\section{1D Random dimer model}\label{sec:1D}
For a 1D quantum system with uncorrelated Anderson disorder, the wavefunction is always localized at any energy and behaves asymptotically as
$\psi_E \left(z\right) \sim e^{-z/\xi\left(E\right)}$, where the localization length, $\xi\left(E\right)$, generally depends on the energy.
In cold atom systems, the density profile is a natural quantity to measure, so the localization length can be measured as a function of
disorder \cite{Aspect08,Inguscio08}, although in practice all measurements are averages over energy due to the distribution of the initial
wavepacket \cite{SanchezPalencia07,Aspect11}.  However, a classical particle in a disordered potential can also be localized simply by random
potential peaks higher than the particle energy.  The localization length of a classical particle increases with energy, as a larger region becomes
classically accessible, but the same is true for a quantum particle.  Therefore, without sufficient knowledge of the potential landscape and the
particle energy, it is difficult to determine whether or not the localization seen in a given cold atom experiment is a quantum effect \cite{Aspect11b}.

However, there are some correlated forms of disorder for which the 1D Anderson localization length does not depend monotonically on energy.
The most famous of these is the random dimer model \cite{DWP},
\be
\label{Hrdm}
H=\sum_{n=1}^{L}\left[\epsilon_{n}c_{n}^{\dagger}c_{n} + t \left(c_{n}^{\dagger} c_{n+1} + c_{n+1}^{\dagger}c_{n}\right)\right],
\ee
where $L$ is the length of the chain, $t$ is the hopping parameter, and $\epsilon_{n} = \pm \epsilon$ is the random on-site potential.
In this model the binary disorder has short-range correlations such that equal energy sites always occur in pairs.

Although uncorrelated disorder localizes all 1D states to some finite length, in the random dimer model the dimer correlation allows a pair
of transmission resonances at critical energies $E_c= \pm \epsilon$ \cite{DasSarma, SS3, DWP, 7SS, SS6}.  There is a boundary value, $\epsilon=t$, such that for larger $\epsilon$
there are no delocalized states \cite{TS, TSO}, but for $\epsilon<t$ (the ``bulk" region), the localization length actually diverges as
$\xi\left(E\right)\sim \left(\frac{E-E_c}{t}\right)^{-2}$.  Right at the boundary, $\epsilon=t$,
\begin{eqnarray}
\label{A}
 \xi(E) =\left\{\begin{array}{l}
4\left(\frac{E-E_c}{t}\right)^{-1}\qquad \qquad \text{ for} \; E\rightarrow E_c^-\\
\sqrt2\left(\frac{E-E_c}{t}\right)^{-1/2}\qquad \;\text{ for} \; E\rightarrow E_c^+
\end{array}
\right.
\end{eqnarray}

\begin{figure}[]
\centerline{\includegraphics[width=.75\columnwidth]{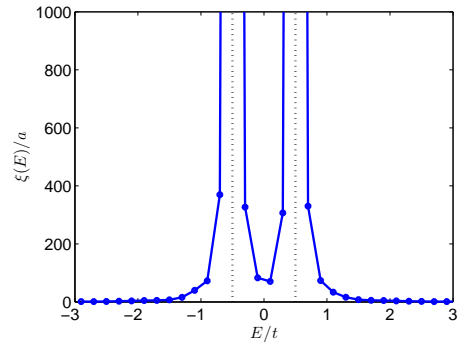}}
\caption{(Color online) Localization length versus energy for random dimer disorder.  Here $\epsilon=t/2$, $a$ is the lattice spacing, and the two values of the on-site energy occur with equal
probability.}
\label{fig:onelegloclength}
\end{figure}

Thus the localization length has peaks, as shown in Fig.~\ref{fig:onelegloclength}, and we see that for correlated disorder the localization
length can actually \emph{decrease} with energy in some cases.  Clearly this is a purely quantum signature associated with true Anderson localization
as opposed to classical localization, as noted in Ref.~\cite{Aspect11b} where certain long-range disorder correlations are predicted to give rise to a
qualitatively similar phenomenon.  However, in the case of Ref.~\cite{Aspect11b} it is not easy to directly measure localization length as a function of
energy because the atoms have a broad initial distribution of energies \cite{Aspect11}.  For the random dimer model, though, the existence of delocalization
points allows the anomalous behavior of the correlation length to be inferred easily: After releasing the atoms from the initial confinement, one simply
waits long enough that the atoms with energies near the critical delocalization energies have propagated out of the lattice, so that the final momentum
distribution for the remaining atoms has dips which will be evident in a time-of-flight image.

\begin{figure}[]
\includegraphics[width=\columnwidth]{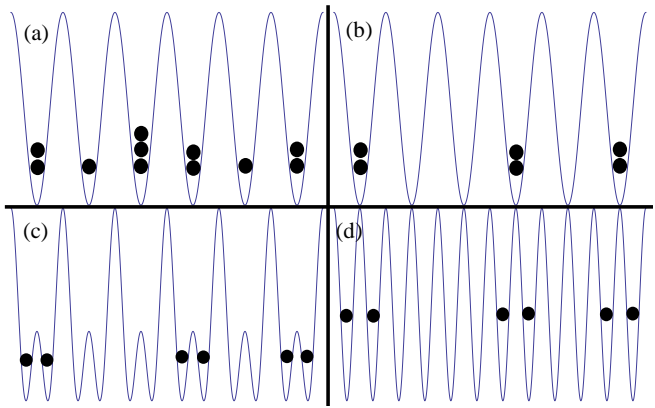}
\caption{(Color online) Scheme to realize random dimer disorder.  (a) Quench to random static distribution.  (b) Purify to empty or doubly occupied sites.  (c)  Split wells
adiabatically.  (d) Random dimer disorder.}\label{fig:oneleg}
\end{figure}

We now turn to the details of realization and observation of the above physics in optical lattice experiments.
Disordered potentials can be realized in a cold atom system either by directly disordering the optical trap \cite{Aspect08,Inguscio08} or by loading a
regular lattice with one mobile atomic species and one immobile atomic species that serves as a pinned impurity distribution for the mobile atoms
\cite{Castin05,Roscilde10,Vignolo10}.  The controllability of nearly all aspects of optically trapped cold atomic systems allows several complementary
 schemes to realize disordered potentials.  Below we outline one possibility, sketched in Fig.~\ref{fig:oneleg}.

To implement 1D random dimer model disorder, one could pin impurity atoms in a 1D optical lattice.  Standard techniques \cite{Thalhammer06,Rempe06,Winkler06}
allow selective removal of atoms such that only empty or doubly-occupied sites remain, with a random distribution.  Each well is then split by adiabatically
turning off the original lattice while turning on a lattice with half its wavelength \cite{Porto06}. Randomly distributed pairs of singly occupied sites and
pairs of empty sites result if the impurity atoms are repulsively interacting or if a second round of purification is performed to remove doubly occupied sites.
(A similar procedure has recently been proposed to realize the dual random dimer model \cite{Vignolo10}, where sites with $+\epsilon$ never occur in pairs, but
that proposal introduces unwanted additional correlations.)  Errors in the disorder initialization may result in additional uncorrelated disorder with a mean
free path $\ell$.  The effect of these errors will be to superimpose another overall localization envelope on the wavefunction, $e^{-z/\ell}$.  We will see that
this is not a handicap, as long as $\ell$ is not small compared to the system size.

\begin{figure}[]
\centerline{\includegraphics[width=\columnwidth]{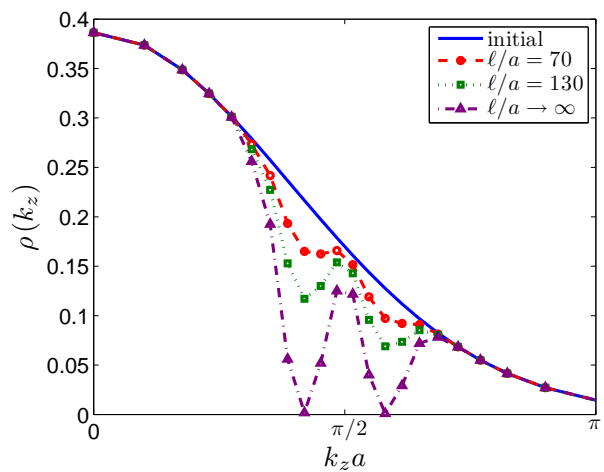}}
\caption{(Color online) Momentum distribution along the z-axis of atoms in random dimer disorder with $\epsilon=t/2$.  The initial momentum distribution is taken here to
be gaussian.  The distribution measured via time-of-flight after the delocalized atoms have left the system is shown in the presence of additional non-dimer
impurities with mean free path $\ell$. The system size is taken to be 100 sites.}
\label{fig:dips}
\end{figure}

Once the disordered potential is produced, the mobile atoms -- bosons with intraspecies scattering length tuned to zero -- can be loaded into the lattice in
the presence of an additional tightly confining potential, as in Refs.~\cite{Aspect08,Inguscio08}.  The final distribution of crystal momenta for the localized
atoms to be imaged is
\begin{multline}
\rho_f\left(K_z\right)= \rho_i\left(K_z\right) \int_{-L/2}^{L/2} dz |\psi_{K_z} \left(z\right)|^2
\\
= \rho_i\left(K_z\right) \int dE \,\delta\!\left(E-2 t \cos{K_z a} \right)
\\
\times \int_{-L/2}^{L/2} dz \left(1/\xi\left(E\right) + 1/\ell\right) e^{-2z\left(1/\xi\left(E\right) + 1/\ell\right)}
\\
= \left(1-e^{-L\left(1/\xi\left(2 t \cos{K_z a} \right) + 1/\ell\right)}\right) \rho_i\left(K_z\right).
\end{multline}
Here $\rho_i\left(K_z\right)$ is the initial momentum distribution, and we have assumed only the first band is populated.  Adiabatically turning off the lattice maps quasi-momentum to real momentum \cite{Winkler06,Greiner01}
and taking a time-of-flight image of these atoms (after turning off interaction with impurity atoms, if applicable) yields a momentum distribution with marked
dips at momenta
\be
k_z = a^{-1} \arccos\left(\pm \epsilon/2t\right),
\ee
corresponding to the delocalization energies.  We show an example of this in Fig.~\ref{fig:dips} for an initial gaussian momentum distribution.  Note that
this signature is somewhat suppressed by any non-dimer impurities present, but is still visible as long as the system size is not much larger than the mean
free path of the non-dimer impurities.  The dips also become narrower for larger systems.  These dips are an unmistakeable signature of quantum
localization-delocalization points.

\section{N-leg anisotropically-correlated disorder}\label{sec:Nleg}
We now generalize to an Anderson disordered system on an $N$-leg optical lattice, which may be thought of as a type of intermediate system between 1D and 2D.
We will again examine the role of correlations in the disorder potential. We consider an extreme case of anisotropic correlations with short-range dimer correlations
along each leg and infinite-range correlations between legs.

The Hamiltonian for this system is
\begin{multline}
\label{H}
H=\sum_{n=1}^{L}\sum_{i=1}^N\bigg\{\epsilon_{n,i}c_{n,i}^{\dagger}c_{n,i} + t (c_{n,i}^{\dagger} c_{n+1,i} + c_{n+1,i}^{\dagger}c_{n,i})
\\
+t_{\perp} (c_{n,i}^{\dagger}c_{n,i+1} + c_{n,i+1}^{\dagger}c_{n,i}) \bigg\}.
\end{multline}
Here $t$ and $t_{\perp}$ are constant intraleg and interleg hopping parameters, respectively, $L$ is the number of sites in each leg, and $N$ is the number of
legs in the ladder.  Energy $\epsilon_{n,i} = \pm \epsilon$ is the random on-site potential.
Denoting by $\psi_{n,i}$ the single-particle wave function at site $n$ of the $i$-th leg and defining
the two-column wavefunction $\Psi_n= (\psi_{n,1}, \dots, \psi_{n,N}, \psi_{n-1,1}, \dots, \psi_{n-1,N})$,
one can write the solution of the Schroedinger equation $H |\psi\rangle = E |\psi\rangle$, where $|\psi\rangle= \sum_{n,i}\psi_{n,i} |n, i\rangle$, as
$\Psi_{n+1} = \prod_{i=1}^n T_i \Psi_1 $, where $T_i$ is the transfer matrix.

Correlations in the random on-site potential are imposed along the length of the chain such that
equal energy sites occur in pairs, or more generally in ``$M$-mers".  Correlations are imposed between chains in various ways to be discussed below.

\subsection{Vertical stripe model}
The simplest example of cross-leg correlations is a system where $\epsilon_{n,i} = \epsilon_{n}$, i.e., all chains are identical copies of a 1D random dimer potential.  In this case, the transfer matrix is
\bea
\label{tm}
T_n=\left(\frac{E-\epsilon_n}{t} I_N- \frac{t_{\perp}}{t} S \right)\otimes\frac{1+\sigma_3}{2}-i I_N \otimes\sigma_2,
\ena
where $\sigma_{1,2,3}$ are the Pauli matrices, $I_N$ is an $N \times N$ identity matrix, and $S$ is an $N$-dimensional matrix responsible for transport in the
vertical direction with $S_{i,j} = \delta_{i, j\pm 1}$.

\begin{figure}[]
\subfigure[]{\includegraphics[width=.45\columnwidth]{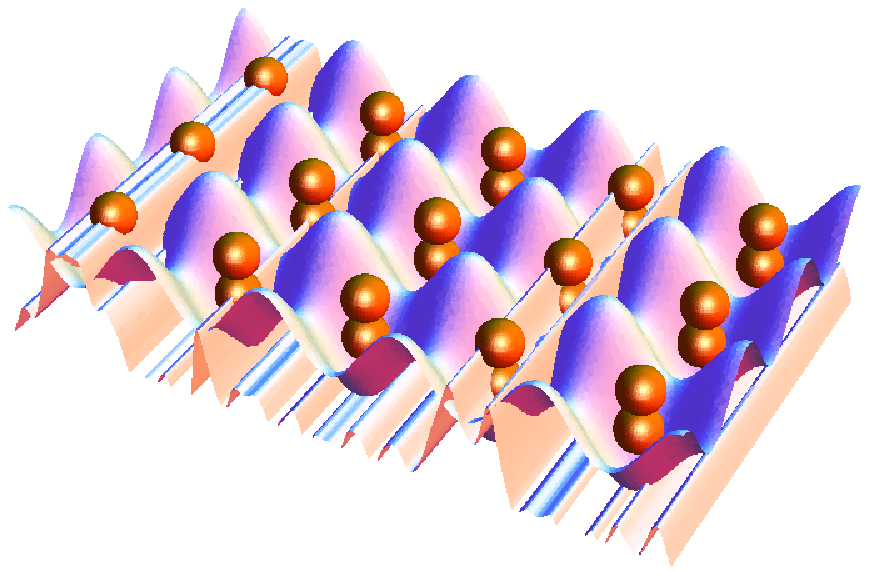}\label{fig:nlegb}}
\subfigure[]{\includegraphics[width=.45\columnwidth]{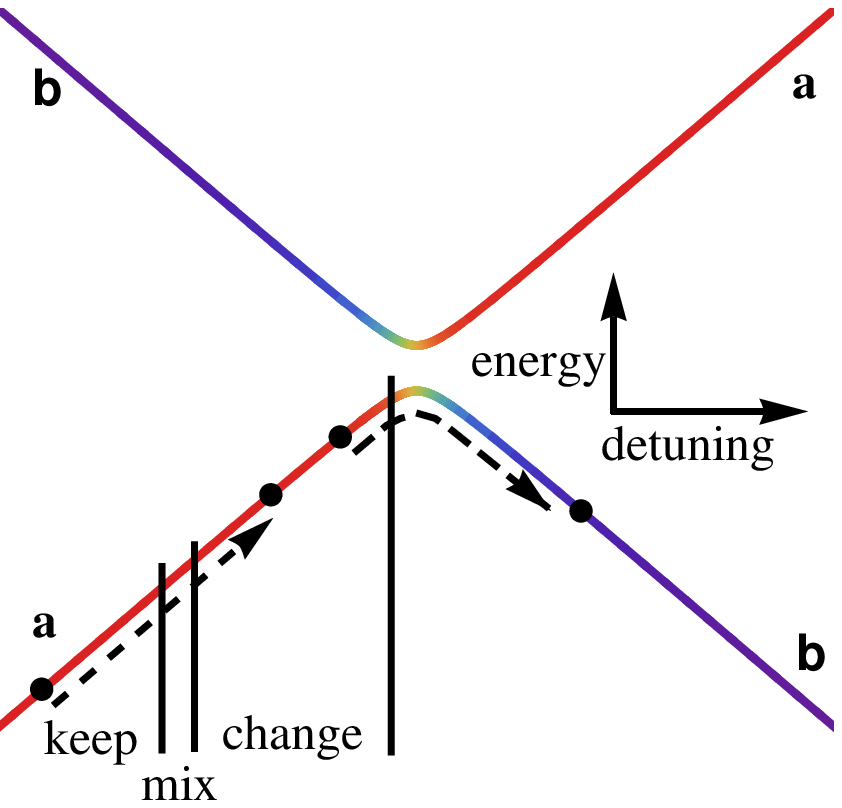}\label{fig:switch}}
\subfigure[]{\includegraphics[width=.45\columnwidth]{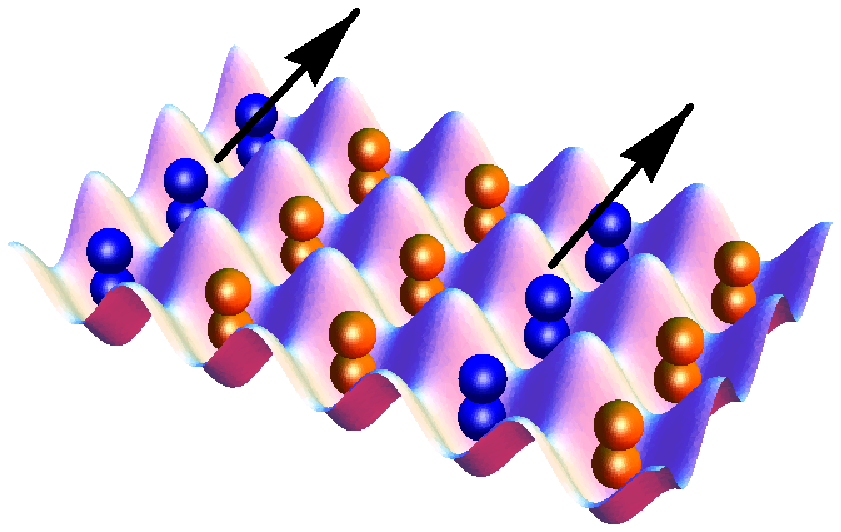}\label{fig:nlegc}}
\subfigure[]{\includegraphics[width=.45\columnwidth]{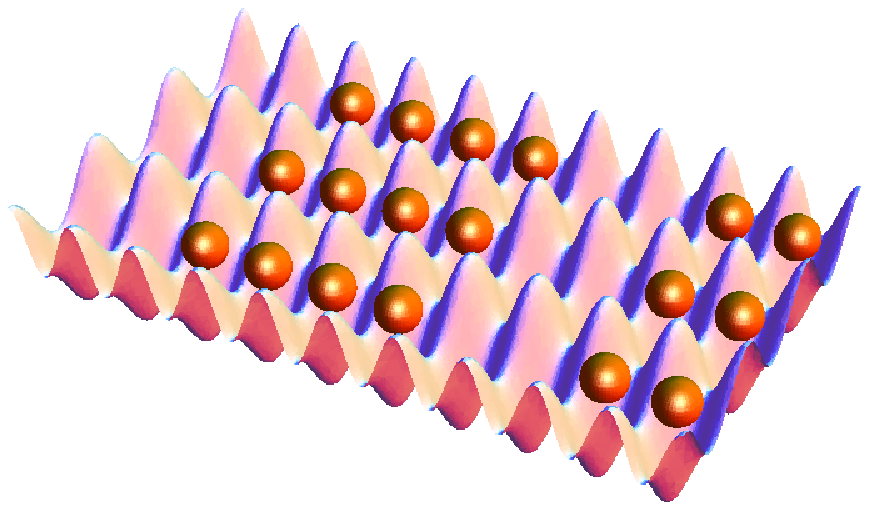}\label{fig:nlegd}}
\caption{(Color online) Scheme to realize $N$-leg random dimer disorder.  (a) Load $n=2$ Mott insulator and apply speckle beam.
(b) Sweep detuning to adiabatically transfer atoms in randomly selected columns from internal state $|a\rangle$ to $|b\rangle$.
State transfer occurs for atoms in sites that are shifted into a window of detunings near resonance by the speckle beam.
The internal state is mixed for a small window of intermediate detunings.  (c)  Remove atoms in state $|b\rangle$ [blue (dark gray)].  (d)  Split wells adiabatically, resulting in dimer disorder.}\label{fig:nleg}
\end{figure}

One can realize this system as in the 1D case, except that now one must initialize randomly distributed \emph{stripes} of empty sites and doubly occupied sites.
 This could be done by loading the lattice with two atoms on every site \cite{Rempe06} and temporarily turning on a optical speckle potential
that is disordered along the length of the chains but constant across the legs, as in Ref.~\cite{Aspect08} (see Fig.~\ref{fig:nlegb}).
The speckle beam introduces a differential light shift so that each cross-leg column of sites has an associated random atomic transition
frequency.  By slowly sweeping the frequency of an external microwave field over an appropriate range, all atoms whose internal transition
frequency is shifted into a given range by the speckle will be adiabatically transferred into a different internal state, as in Ref.~\cite{Kuhr11}
(see Fig.~\ref{fig:switch}).  Upon adiabatically turning off the microwave field and the speckle, one is left with randomly distributed stripes of
sites with atoms in a different internal state.  These atoms can be removed with a blast pulse.  Splitting each well in the intrachain direction as
before results in the desired disorder.  We sketch the process in Fig.~\ref{fig:nleg}.

Note that, for a continuous distribution of random speckle shifts, there will always be a fraction of columns for which the adiabatic
sweep results in a mixed state, as shown in Fig.~\ref{fig:switch}.  This leads to imperfections in the disorder pattern after the projective
blast pulse.  If one sweeps the microwave field 60 kHz over 20 ms with pulse parameters as in Ref.~\cite{Kuhr11}, the detuning range that
gets transferred to a mixed state is roughly 2 kHz wide.  Assuming the random light shifts have a frequency range similar to that of the detuning sweep
(corresponding to a 2 mW speckle beam \cite{Clement06} in this case), and accounting for errors from nonadiabatic transitions as well, one
obtains an average distance between errors in the random dimer disorder of $\ell \sim 70$ lattice spacings.  However, $\ell$ can be greatly
increased by using a slower sweep.

Since the potential is separable in the inter- and intrachain directions, one simply obtains the same localization length as a function of
energy along the chain as in the 1D case above (see App.~\ref{app:A}), and the same dips in the momentum profile.  The only difference is that there will be $N$
times as many delocalized states, due to the $N$ possible transverse modes.

\subsection{Diagonal stripe model}
\begin{figure}[]
\center
\centerline{\includegraphics[width=.9\columnwidth,angle=0,clip]{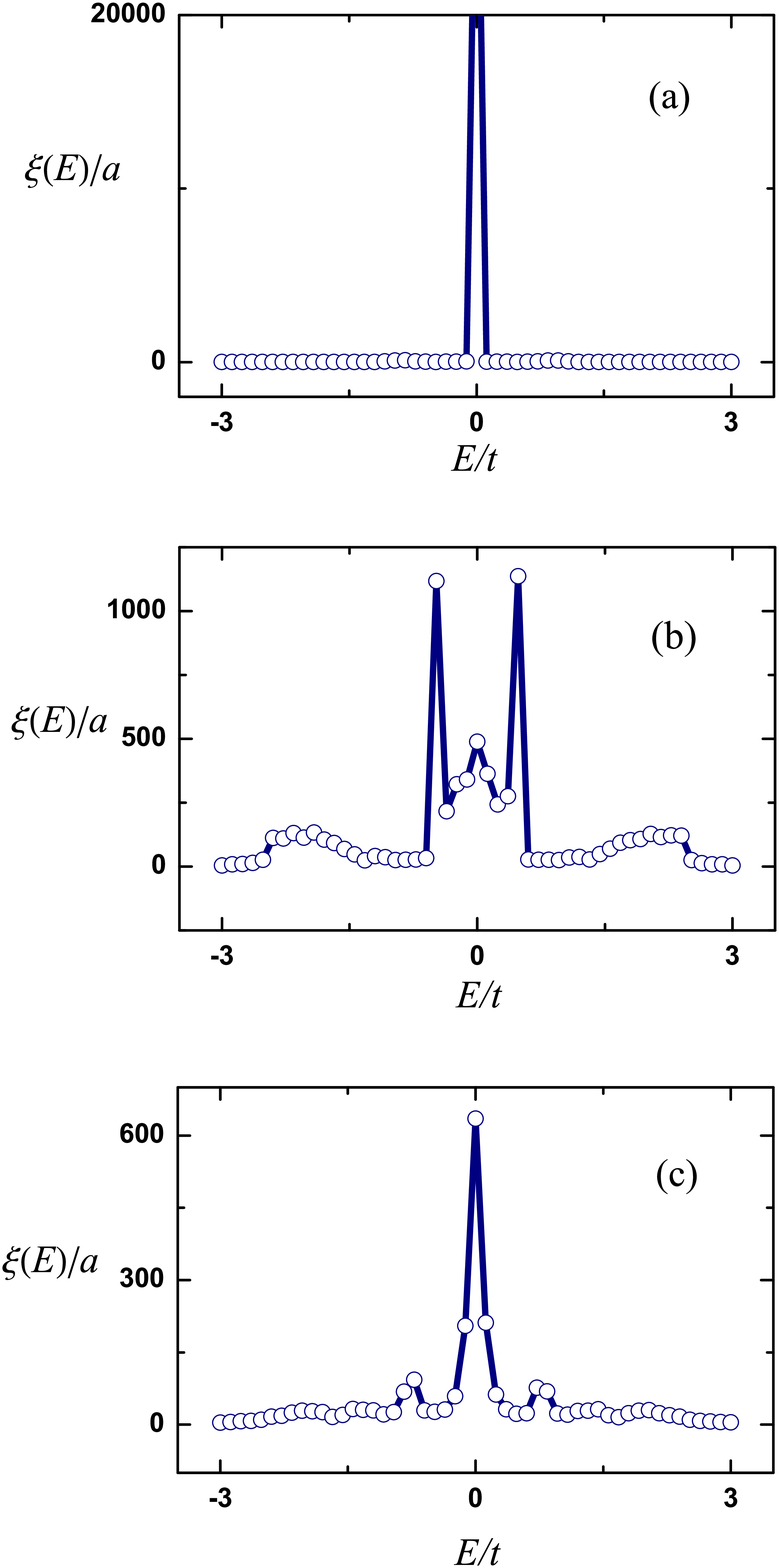}}
\caption{(Color online) Localization lengths for two-leg dimer ($N=2$, $M=2$) diagonal stripe model at: (a)
$\epsilon=t/2$ and $t_{\perp}=t/2$; (b) $\epsilon=t/2$ and $t_\perp=t$; (c) $\epsilon=3t/4$ and $t_\perp=t$. Numerical calculation was done for a system with 8000 sites in the horizontal direction.
}
\label{fig:diagonal}
\end{figure}
Clearly, we must break the separability of the disorder potential in the intra- and interchain directions to obtain different behavior.  The simplest way to do that is to rotate the propagation axis of the speckle beam in the 2D plane such that it is no longer perpendicular to the chains.  For instance, if the same procedure as above is performed with the
speckle beam pointed at a $45^{\circ}$ angle to the legs in the plane, then each leg will have identical dimer sequences (except for at the boundaries), but the
sequence on each leg will be shifted by two lattice sites relative to the leg above it, such that $\epsilon_{n,i}=\epsilon_{n-2,i+1},\; i=1,\cdots N-1$.
Here, the transfer matrix is given by

\bea
\label{tm}
T_n=
\left(\frac{E}{t}I_N-Y_{n,N} - \frac{t_{\perp}}{t} S \right)
\otimes\frac{1+\sigma_3}{2}
-i I_N \otimes \sigma_2,
\ena
where $Y_{n,N}$ is an $N$-dimensional diagonal matrix with entries: $(Y_{n,N})_{jj}=\epsilon_{n+2(j-1),1}\delta_{jj}$.

We have used this form of the transfer matrix to calculate the localization length of a two-leg dimer system numerically; see Fig.~\ref{fig:diagonal}. In Fig.~\ref{fig:diagonal}(a)  we see the presence of delocalization transition in this model at zero-energy when $\epsilon=t_\perp$. This point is special in the sense that it corresponds to the case when the on site chemical potential is sufficiently large to guarantee for a particle to hop freely in the vertical direction. This
effectively restores the translational invariance in the vertical direction and makes the model analogous to the vertical stripe model. Supporting evidence is that the critical index corresponding to this transition is $\nu=2$, which coincides with that of 1D dimer model in the "bulk" region.
For other values of model parameters the single particle states remain localized; see Fig.~\ref{fig:diagonal}(b) and (c) for two
particular choices of model parameters.

\subsection{Staggered vertical stripe model}
\begin{figure}[]
\center
\centerline{\includegraphics[width=.9\columnwidth,angle=0,clip]{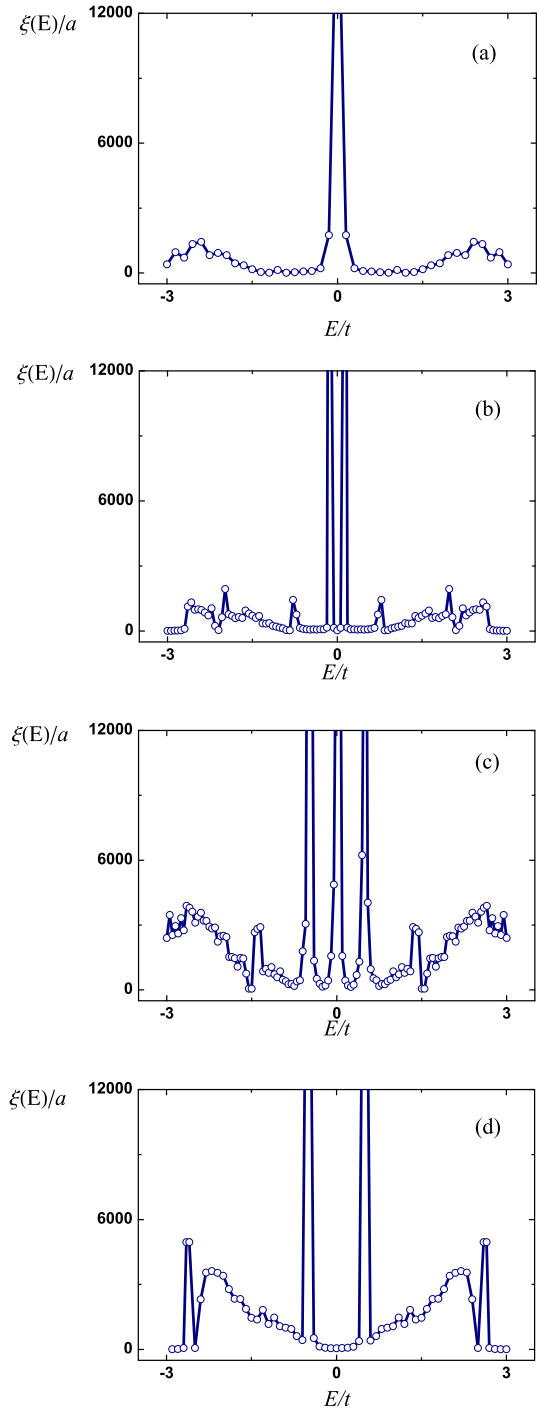}}
\caption{(Color online) Localization lengths of staggered vertical stripe models for $\epsilon=t/2$.
(a) Two-leg dimer model ($N=2$, $M=2$) showing $E_c \simeq 0$, (b) two-leg trimer model ($N=2$, $M=3$) showing $E_c \simeq \pm 0.12t$, (c) three-leg dimer model ($N=3$, $M=2$) showing $E_c \simeq \pm 0.5t$.
(d) two-leg trimer model ($N=2$, $M=3$) showing $E_c \simeq \pm 0.5t$.}
\label{fig:LL1}
\end{figure}
Another interesting disorder pattern that is not separable is dimer-correlated disorder in each chain with long-range interchain correlations such that each
chain is the mirror image of the one above it, i.e., the disorder potential is staggered along the interchain links.  The Hamiltonian of the model is defined by Eq.~\eqref{H} where the site energies $\epsilon_{n,i}=(-1)^i \epsilon_n$ along the vertical direction $i=1,..N$ depend on a single random parameter $\epsilon_n=\pm \epsilon$.  The transfer matrix of the model reads
\bea
\label{tm1}
T_n=\left(\frac{E I_N-\epsilon J_N}{t} - \frac{t_{\perp}}{t} S \right)\otimes\frac{1+\sigma_3}{2}-i I_N \otimes\sigma_2,
\ena
where $I_N $ is identity matrix, while $(J_N)_{jj\prime}=(-1)^j \delta_{jj^\prime}$.

In Fig.~\ref{fig:LL1} we present data for localization lengths
in a staggered model for two-leg dimer (a) and  trimer (b),  and three-leg dimer (c)
and trimer (d) models, respectively. All numerical simulations here have been done for a chain of 6000 $M$-mers.
  We see here a very rich collection of true delocalized states.
It is important to notice that the staggered model can not be reduced to a set of $N$ one-leg models as in the plain vertical stripe case.

\begin{figure}[]
\center
\centerline{\includegraphics[width=.9\columnwidth,angle=0,clip]{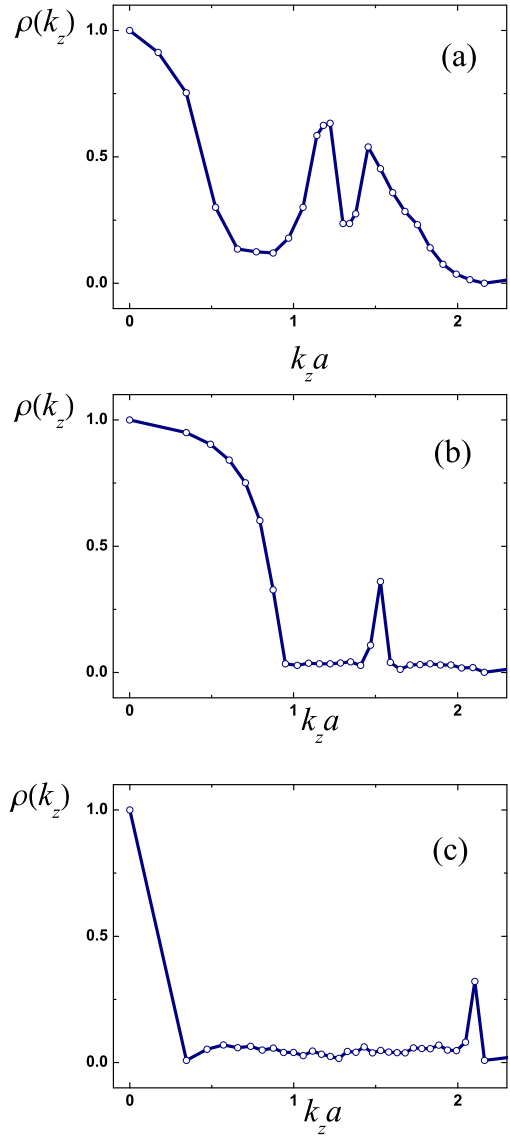}}
\caption{(Color online) Momentum distribution after delocalized states have left the sample
for N=2, M=2  (a);
N=2, M=3 (b); and N=3, M=2 (c) cases, respectively. The initial distribution is the same as in Fig.~\ref{fig:dips}.}
\label{fig:Smomdist}
\end{figure}

Fig.~\ref{fig:LL1} clearly shows the presence of delocalization of the states in all  cases (a-d), however another interesting property
of the staggered model is evident here. Namely, we see a region of well-extended states around $E \sim 1.5 t - 3 t $.
Their localization lengths are comparable with the length of the chain under consideration.
This fact produces large regions of dips in a momentum distribution of the states (see Fig.~\ref{fig:Smomdist}).

Fig.~\ref{fig:Smomdist} represents distributions of horizontal (i.e., interleg) momentum after delocalized and extended states have left a sample with 100 sites.
We take an initially Gaussian momentum distribution centered at zero.

From Appendix~\ref{app:B}, we have that the connection of the delocalization energy $E_c$ and the delocalization momentum $p_c$
is defined by the expression
\begin{eqnarray}
\label{con}
p_c=
  \arccos\left[\frac{E_c \pm \sqrt{t^2+\epsilon^2}}{2t}\right],
\end{eqnarray}
for $N=2$, and
\begin{eqnarray}
\label{con1}
p_c=
\left\{
\begin{array}{cc}
\arccos[\frac{E_c \pm \sqrt{2 t^2+\epsilon^2}}{2t}]&   \\
  \arccos[\frac{E_c -\epsilon}{2t}]             &
\end{array}
\right.
\end{eqnarray}
for $N=3$.
It is important to note here that the spectrum depends on the number of legs, $N$, in a nontrivial way.  This implies $N$-dependence of observables in general and of the momentum distribution in particular (see Fig.~\ref{fig:Smomdist}).

\begin{figure}[h!]
\subfigure[]{\includegraphics[width=.45\columnwidth]{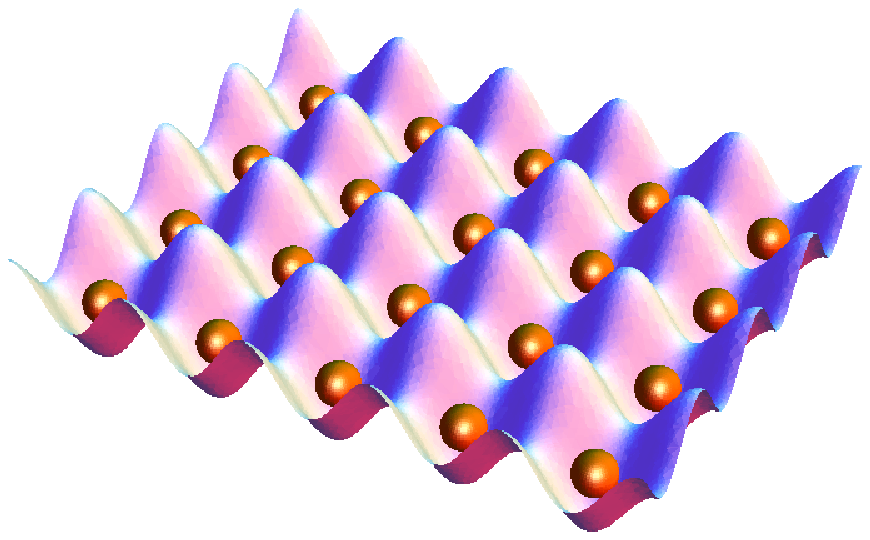}\label{fig:staga}}
\subfigure[]{\includegraphics[width=.45\columnwidth]{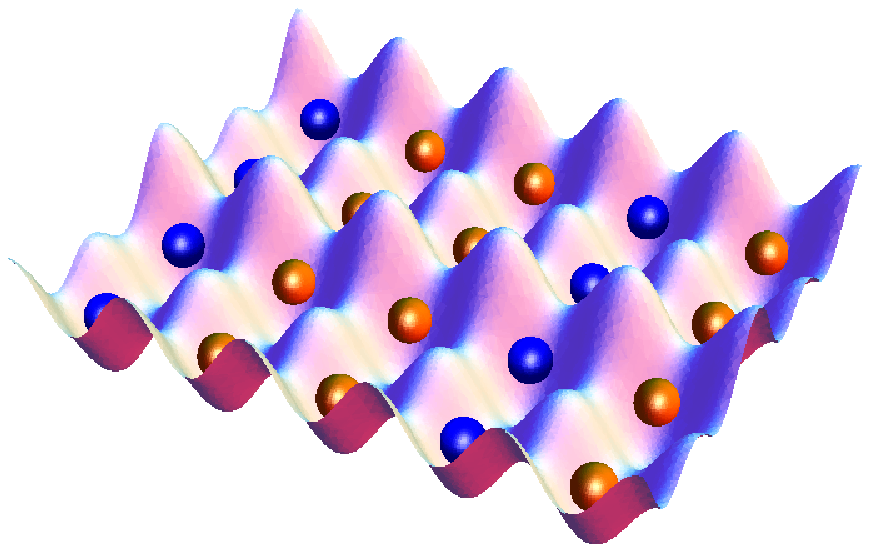}\label{fig:stagb}}
\subfigure[]{\includegraphics[width=.45\columnwidth]{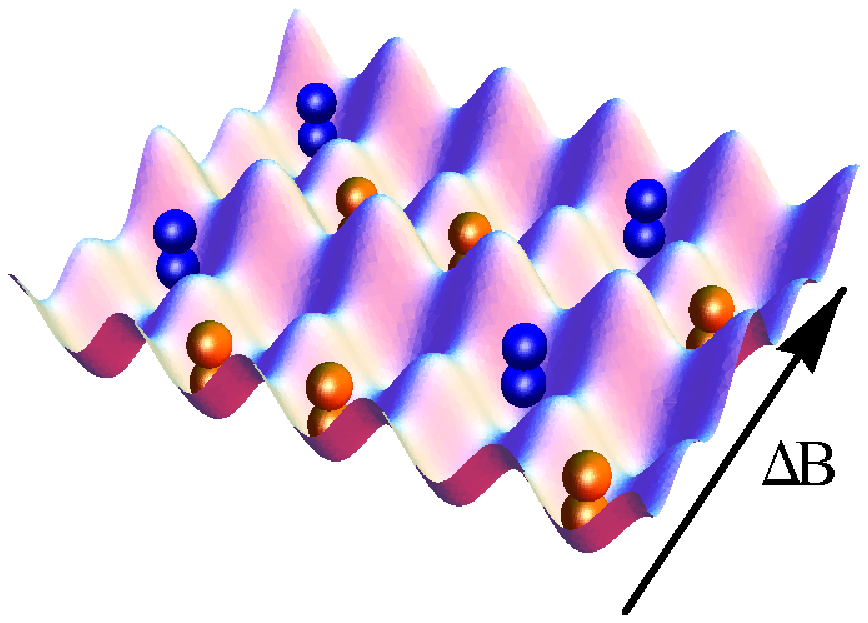}\label{fig:stagc}}
\subfigure[]{\includegraphics[width=.45\columnwidth]{staggeredEc.eps}\label{fig:stagd}}
\caption{(Color online) Scheme to realize staggered $N$-leg dimer disorder.  (a) Load $n=1$ Mott insulator.  (b) Flip internal state of atoms in randomly
selected columns [blue (dark gray)] as in Fig.~\ref{fig:nleg} and lower every other interleg barrier.  (c)  Apply magnetic field gradient to obtain state-dependent tilt.  (d)
Restore barriers, flip back internal states, and split wells adiabatically.}\label{fig:stag}
\end{figure}

A staggered disorder potential can be realized experimentally by loading the lattice into a $n=1$ Mott insulator state, then using the speckle beam to select columns of sites to flip to
a different internal state, as in the vertical stripe case discussed above.  However, here one must specify the two relevant internal states to have opposite Zeeman shifts.
Then, deforming the lattice to a two-period superlattice, ramping up the tunneling, applying a magnetic field gradient, and ramping the tunneling back down
results in the atoms being shifted according to their internal state, as shown in Fig.~\ref{fig:stag}.  This is somewhat similar to the procedure used to
initialize a Neel state in Ref.~\cite{Trotsky06}.

More straightforward methods of realizing the short-range correlated disorder are also available, though they require sophisticated single-site control.
In that case, one can create completely arbitrary disorder patterns.  One possibility is to use a custom holographic mask as in Ref.~\cite{Greiner09} to
generate a two-dimensional (2D) lattice potential with the desired disorder pattern.  Another is to use a regular 2D optical lattice, but to etch the desired
disorder pattern onto a regular array of impurity atoms with single-site resolution \cite{Ott09,Kuhr11}.

\section{Conclusions}\label{sec:conclusions}
We discussed a class of quasi-$1D$ disordered models with correlations between random variables, which,
 depending on the symmetries of the model which we explicitly identify and discuss, may or may not exhibit localization-delocalization transitions.
 We establish using a few examples how such models and their observable characteristics (such as momentum distribution and single particle spectrum) strongly depend on symmetry properties of the effective hopping Hamiltonian, which, in turn, classify the models accordingly.

We have shown that cold atoms trapped in an optical lattice can realize the well-known one-dimensional random dimer model, which exhibits critical delocalization energies, and proposed an experimental procedure to observe the delocalization via time-of-flight imaging.  We have also generalized the random dimer model to $N$-leg systems with highly
anisotropic correlated disorder and found new transmission resonances induced by the transverse degree of freedom.  We have calculated the localization length as a function of energy and identified the critical energies.  These resonances, both in the 1D and $N$-leg systems, present unique and unambiguous signatures of the quantum delocalization energies.  Furthermore, the $N$-leg case provides an exciting prospect for experiments to explore correlation-induced delocalization outside of the strictly 1D limit with the current state of the art.  Also, any experimental observation of critical energies separating localized and delocalized states in well-designed disordered optical lattices along the line of our suggestions in this work will immediately establish the occurrence of the quantum localization phenomena in cold atomic sysems.

We thank T. Porto and S. Rolston for helpful discussion.  This work supported by AFOSR-MURI, ARO-MURI, DARPA-OLE, and NSF-JQI-PFC.

\appendix

\section{}\label{app:A}
Localization properties of the system can be investigated by calculating Lyapunov exponents. According to Oceledec's theorem~\cite{O},
the $\frac{1}{L}$ power of the product of $L$ random transfer matrices has a limiting value
\bea
\label{OS}
\Lambda = \lim_{L\rightarrow\infty} \Bigg[\prod_{n=1}^L T_n \prod_{n=L}^1 T^{\dagger}_n\Bigg]^{1/2 L}.
\ena
The closest to unity eigenvalue of $\Lambda$ can be represented as $e^{\lambda(E)}$, where $\lambda(E)$ is the Lyapunov exponent and the localization length is simply its inverse,
$\xi(E) = 1/\lambda(E)$.  A delocalization transition occurs at energies $E_c$ where the localization length diverges.
We analytically calculate the critical energies, $E_c$, for $N$-leg $M$-mer disorder in two steps. First we reduce the $N$-leg problem to an effective
one-leg problem. Then we solve the problem for arbitrary $M$-mers in a one-leg chain.

The eigenvalues of $t_{\perp}S$ are
\bea
\label{shift}
E_{N,k}=\Bigg\{\begin{array}{ccc}
         2 t_{\perp} \cos\left(\frac{\pi k}{N+1}\right),\;\;& k=1, \dots, N, \;\;& N>2\nn\\
         t_{\perp} \cos\left(\pi k\right), \;\;& k=1,2 \qquad \qquad & N=2  \\
         0   \qquad \qquad & \qquad \qquad \qquad \qquad & N=1
        \end{array},
\ena
so the whole transfer matrix can be recast in the form of $N$ independent 1D transfer matrices corresponding to different transverse modes,
$T_n = \bigoplus_k T_n^{(k)}$, where
\bea
\label{tmd}
T_n^{(k)}=\Big(\frac{E-\epsilon_n- E_{N,k}}{t} \Big)\frac{1+\sigma_3}{2}-i \sigma_2.
\ena
Therefore, the critical energies, $E_c$, of the ladder model correspond to those of a single decoupled chain plus the transverse mode energy.

The $M$-mers are now treated by defining a new basic transfer matrix, which is the $M$th
power of the expression \eqref{tmd}, $\big[T^{(k)}_{n}\big]^M$, with randomly distributed site energies $\epsilon_n$.
At a delocalization energy, $E_c$, the transfer matrix has an eigenvalue being $\pm 1$ \cite{DasSarma} (in units of $t$).  Here this implies that for one of the values of $k$,
\bea\label{deloc}
\big[T^{(k)}_{n}\big]^M = \pm I_2
\ena

To solve this equation it is convenient to express the transfer matrix \eqref{tmd} in terms of trigonometric functions,
\bea
\label{SU}
T_n^{(k)}= e^{i {\bm \omega}\cdot{\bm \sigma} }= I_2 \cos\omega +i \frac{{\bm \omega}\cdot{\bm \sigma}}{\omega} \sin\omega,
\ena
where $\omega=\arccos\left(\frac{E-E_{N,k}-\epsilon_n}{2 t}\right)$ and $\bm{\omega}=\left(0, -\frac{\omega}{\sin\omega}, -i \omega \cot\omega\right)$.

Then
$\big[T^{(k)}_{n}\big]^M= I_2 \cos\left(M \omega\right)+i \frac{{\bm \omega \cdot \sigma}}{\omega} \sin\left(M \omega\right)$. Therefore,
the delocalization condition is satisfied for
$\omega=\pi q/ M,\; q=1, \dots, 2 M$, and the critical energies are
\begin{multline}
\label{sol}
E_c= 2 t \cos{\frac{\pi q}{M}} + E_{N,k} \pm \epsilon,
\\
q=1,\dots, 2 M-1;\; q\neq M,\; \text{and} \;  k=1,\dots,N.
\end{multline}
Note that the trivial solutions $\omega=0, \pi, 2 \pi $, corresponding to transfer matrix $T_n\equiv \pm I_{2N}$, should not be considered.
Below we write explicitly the critical points for dimers, trimers, and tetramers:
\bea
\label{simsol}
M=2:&& E_c=  E_{N,k} \pm \epsilon;\nn\\
M=3: && E_c=  E_{N,k} \pm t_{\perp} \pm \epsilon;  \\
M=4: &&  E_c=  E_{N,k} \pm \sqrt{2} t_{\perp} \pm \epsilon,\;\;  E_{N,k}\pm \epsilon.\nn
\ena

Generally speaking, Eq.~\eqref{deloc} is a sufficient condition for having a delocalization transition but it is not a necessary one. Therefore, we have
checked numerically for up to $N=7$ legs and for correlations of up to $M=5$ sites that Eq.~\eqref{sol} captures all the critical
points.

Away from the critical points, the localization length must be calculated numerically from Eq.~\eqref{OS}. Fig.~\ref{fig:onelegloclength} shows some numerical results for localization lengths in a chain of $L=5000$ sites. The critical energies obtained are in perfect agreement with the analytical result of Eq.~\eqref{simsol}.

\section{}\label{app:B}
In this Appendix we will calculate the spectrum of a particle in the staggered vertical stripe model (class $Z_2$).

In order to obtain the relation between the energy and the momentum of the free on-shell particle in a staggered disordered model,
one needs to calculate the dispersion of the free particle in the staggered but {\em non-random} model
defined by the Hamiltonian (\ref{H}) with $\epsilon_{n,i}=(-1)^i \epsilon$ (with non-random $\epsilon$).
The reason for this is the following. Theoretically we know that the particles in a disordered chain are either localized ($\xi(E)$ is finite) or delocalized ($\xi (E_c)\rightarrow\infty$). Experimentally the localized on-shell particles are being "detected" by measuring their momentum distribution profile. At the time when the measurement is done the energy of particles is linked to the momentum in a way which has no information about disorder and its pre-history, and therefore the spectrum of the detected particle precisely coincides with the spectrum of the free particle in a non-random model.

  As the first step we Fourier transform the Hamiltonian with no disorder in the horizontal direction and introduce dimensionless momenta, $p_z$. Then the Hamiltonian for the two leg dimer system, $N=2$,
(eigenfunctions of which are $\psi_{i,p}$, $i=1,2$) becomes
\bea
\label{dispersion}
H= \left(
\begin{array}{cc}
\epsilon& t_{\perp}\\
t_{\perp}& -\epsilon
\end{array}
\right) + 2 t \cos[p_z] \hat{1} .
\ena
Diagonalization of this Hamiltonian yields the following dispersion relation
\bea
\label{dispersion2}
E=\pm \sqrt{\epsilon^2+t_{\perp}^2}+ 2 t \cos[p_z]
\ena
which produces Eq.~(\ref{con}).

For three leg system we have the following Hamiltonian in the momentum space
\bea
\label{dispersion3}
H= \left(
\begin{array}{ccc}
\epsilon& t_{\perp}&0\\
t_{\perp}& -\epsilon&t_{\perp}\\
0& t_{\perp}& \epsilon
\end{array}
\right) + 2 t \cos[p_z] \hat{1} .
\ena
Upon diagonalization of Eq.~(\ref{dispersion3}) we arrive to the following spectrum
\bea
\label{dispersion4}
E=\left\{
\begin{array}{cc}
\pm \sqrt{2 t_{\perp}^2+\epsilon^2}+ 2 t \cos[p_z]&     \\
\epsilon   + 2 t \cos[p_z]          &
\end{array}
\right.
\end{eqnarray}
which defines Eq.~(\ref{con1}) of the main text.

\end{document}